\documentclass[letterpaper, 10 pt, conference]{ieeeconf}
\IEEEoverridecommandlockouts

\usepackage{graphicx}
\usepackage{caption}
\usepackage{algpseudocode}
\usepackage{amsmath}
\usepackage{longtable}
\usepackage{adjustbox}

\usepackage{amsmath,amssymb,amsfonts,epsf,epsfig,times}
\usepackage[all]{xy}
\usepackage{graphicx,color}
\usepackage{subfigure}
\usepackage{url}
\usepackage{cite}
\usepackage{tikz}
\usepackage{epstopdf}
\usepackage{algorithm}
\usepackage[]{amssymb}

\def\QED{\mbox{\rule[0pt]{1.5ex}{1.5ex}}}

\newcommand{\OMIT}[1]{}

\makeatletter
\def\BState{\State\hskip-\ALG@thistlm}
\makeatother

\begin{document}
\title{Multi-Sensor Data Pattern Recognition for Multi-Target Localization: A Machine Learning Approach}
\author{Kasthurirengan Suresh, Samuel Silva, Johnathan Votion, and Yongcan Cao
\thanks{The authors are with the Department of Electrical and Computer Engineering, The University of Texas, San Antonio, TX 78249. 
}
\thanks{Corresponding Author: Yongcan Cao (yongcan.cao@utsa.edu)}
}

\markboth{}
         {}

\maketitle

\begin{abstract}
Data-target pairing is an important step towards multi-target localization for the intelligent operation of unmanned systems. Target localization plays a crucial role in numerous applications, such as search, and rescue missions, traffic management and surveillance. The objective of this paper is to present an innovative target location learning approach, where numerous machine learning approaches, including K-means clustering and supported vector machines (SVM), are used to learn the data pattern across a list of spatially distributed sensors. To enable the accurate data association from different sensors for accurate target localization, appropriate data pre-processing is essential, which is then followed by the application of different machine learning algorithms to appropriately group data from different sensors for the accurate localization of multiple targets. Through simulation examples, the performance of these machine learning algorithms is quantified and compared. 
\end{abstract}


\IEEEpeerreviewmaketitle


\section{Introduction}

Conducting surveillance missions using sensor networks is essential for many civilian and military applications, such as disaster response~\cite{GeorgeEtAl10}, border patrol~\cite{OnurEDA07}, force protection~\cite{AhmedCCK15,CasbeerMeierCao13}, combat missions~\cite{BokarevaEtAl06}, and traffic management~\cite{min2011real}. One main task in these missions is to collect data regarding the operational environment and then obtain intelligence information from the data. Because the sensors used to collect data are often spatially distributed, extracting data patterns becomes critical to obtain accurate knowledge about the underlying activities.

The existing work on identifying data patterns from spatially distributed sensors is focused on developing probabilistic reasoning techniques without recognizing the specific data association or data patterns. Existing approaches for multi-target state estimation can be characterized by two features: a data-to-target assignment algorithm, and an algorithm for single target state estimation under pre-existing data-to-target associations. With unknown data association, probabilistic data association (PDA)~\cite{ShalomDaumHuang09} and multiple hypothesis tracking (MHT)~\cite{Blackman04} are two common approaches where dense measurements are available. In the study of traffic patterns, the existing research is focused on estimating traffic density and smart routes~\cite{min2011real} without analyzing the data pattern to obtain better knowledge of traffic information.


The main limitation of the existing research is the lack of work that addresses data patterns in spatially distributed sensors, which is crucial in obtaining accurate modeling for the multi-target localization problem. For example, if two observations/measurements are obtained for a single target, the estimated location of the target can be better calculated if such a data correlation can be identified before estimation. If the correlation is not identified, the target's location cannot be estimated accurately. Although, an incorrect estimation of total target numbers is expected. Existing techniques do not take into consideration the critical data patterns in multi-target localization. Our proposed method leverages machine learning tools to uncover the data patterns in order to provide accurate multi-target localization and state estimation.   


In this paper, we consider the multi-target data pattern recognition problem where a number of targets move along a known road network with numerous sensors placed at unique random known locations. The target's data is recorded as it passes by the sensor's location. It is assumed that the sensors are unable to identify individual targets. The objective of this paper is to develop algorithms that can effectively extract and associate correlated data to the same target. More precisely, we seek to classify all data obtained from spatially distributed sensors into datasets and associate each of these sets to one target. To achieve this objective, we first generate datasets for targets with different motions. Then a pre-processing algorithm is developed that allows the original dataset to be translated into another form that yields classifiable datasets. Numerous machine learning algorithms, including K-means clustering and support vector machines (SVM), as well as their variations, are used to classify the translated dataset. These algorithms are evaluated via numerous simulation studies that suggest some well-behaved algorithms. 

The remainder of this paper is organized as follows. In Section~\ref{sec:pre}, two standard clustering algorithms: K-means and support vector machine (SVM) are reviewed. Section~\ref{sec:PF} gives the problem formulation. Section~\ref{sec:alg} describes the proposed algorithms that classify datasets for different targets. Section~\ref{sec:simu} provides numerous simulations that illustrate the performance of the proposed algorithms. Finally, Section~\ref{sec:con} provides a brief summary and discusses some potential future work.




\section{Preliminaries}\label{sec:pre}

An important form of learning from observations is to conduct classification of datasets associated with these observations. Traditional techniques generally arrange objects or data based on their numerical measure and consequently their similarity/disposition on a feature graph\cite{michalski1983learning}. Among the several techniques in machine learning, K-means and SVM are two popular classification algorithms that will be used in the context of the paper.

\subsection{K-means clustering} \label{sec:Kmeans}
K-means is one of the most popular data processing algorithms. It is a simple unsupervised learning algorithm used to solve the clustering problem.
The K-means method seeks to minimize the average squared distance between points in the same cluster\cite{arthur2007k}.

Let $X = \{x_1,...,x_n\}$ be a set of $n$ points in a $d$-dimensional Euclidean space. Let $k$ denote the number of clusters. The Euclidean distance between two points $x_i$ and $x_j$ is defined by $\left \| x_i-x_j \right \|$, where $\left \| \cdot \right \|$ is the 2-norm. For a subset $Y \subseteq X $ and a point $x$, the distance between them is defined as $d(x,Y) = {min}_{y\in Y}\left \| x-y \right \|$   \cite{bahmani2012scalable}. The \textit{centroid} for any subset $Y \subseteq X $ is denoted as $\Omega$ and is defined as
\begin{align}\label{eq:centroid}
\Omega(Y) = \frac{1}{\left | Y \right |}~\sum_{y\in Y} y.
\end{align}
A cost value is assigned to each centroid and is related to the distance between the position of the centroid and each point in $X$. Let $\mathcal{C} = \{c_1,...,c_k\}$ be a set of $k$ centroid positions.  The \textit{cost} of $Y$ with respect to $\mathcal{C}$ is defined as
\begin{align}\label{eq:cost_k-mens}
\Phi_Y(\mathcal{C}) = \sum_{y\in Y} d^2(y,\mathcal{C}) = \sum_{y\in Y} \min_{i=1,...,k} \left \|y-c_i  \right \|^2.
\end{align}

The K-means algorithm seeks to find a set $\mathcal{C}$ of $k$ centers to minimize $\Phi_X(\mathcal{C})$. The label/cluster to which each data sample belongs is defined as the one to which the distance between $c_i$ and $X$ is smaller than to any $c_j$, $j\neq i$.

The optimal solution to the k-means problem uses \textit{Lloyd's} iteration \cite{ostrovsky2006effectiveness}. To achieve the placement of the centroids that lead to a minimum cost $\Phi_Y(\mathcal{C})$, the centroids' position is updated at each iteration. At the first loop, the $k$ centroids are placed alongside with the data points at random positions. Given the position of these centroids, for each data point $x_i$ we find the nearest centroid $\Omega$. After going through the entire dataset, a cluster $\mathcal{C}_k = \{x_1,...x_u\}$ is formed containing all $x_i$ which its closer cluster is $\mathcal{C}_k$. Having all clusters being formed, the position of the centroid is then recalculated according to \eqref{eq:centroid}. For the next iteration, the current centroid position is used to form new $\mathcal{C}_k$ respecting the fact that $x_i \in \mathcal{C}_k$ if and only if, its distance $d(x_i,\mathcal{C}_k) \leq d(x_i,\mathcal{C}_u) ~ \forall u$. Such an iterations continues until no significant change is observed on the centroid positions. Given the simplicity of the algorithm, the convergence is expected to be achieved in a few iterations.

It has been shown in \cite{arthur2007k} that with proper definition of initial set $\mathcal{C}$ the accuracy and convergence can be drastically improved. Section~\ref{sec:alg} will describe this idea further.



\subsection{Support Vector Machines (SVM) } \label{sec:SVM}
Support Vector Machines is a supervised learning method used commonly for classification and regression. It consists of minimizing the empirical classification error while maximizing the geometric margin in classification. This approach differs from most of the commonly used machine learning algorithms because it not only aims at simply minimizing the empirical classification error, but also increasing the dimensional feature space to optimize the classification function\cite{yeo2009can}. 

The SVM implements an idea where the input vectors are mapped into a high-dimensional feature space through an \textit{a priori} nonlinear mapping, identified later as `kernel'. In this high-dimensional space a linear surface can be chosen in order to ensure generalization of the network \cite{cortes1995support}.

Given a dataset $D$ with $N$ samples, consisting of elements in the pattern ${(x_j,y_j)^N_{j=1}}$, $\textbf{x} \in R^d$ is the $j^{th}$ sample and $y_j \in (0,1)$ is the corresponding class label. Given these two classes, the objective is to design a hyperplane able to divide them with the largest margin between classes possible \cite{duda2012pattern}. Nonlinear SVM's are often used when a data is not linearly separable in the input space $R^d$. In this case the nonlinear SVM maps the feature vector $\textbf{x} \in R^d$ into a high (up to infinity) dimensional Euclidean space, $H$, by using the nonlinear function $\Phi : R^d \mapsto H$. For this nonlinear relation, the decision boundary for a 2 class problem, takes the following form as
\begin{align}\label{eq:hyperplane}
\centering
\textbf{w}\cdot \Phi(\textbf{x}) + b =0. 
\end{align}
And we obtain $H(\textbf{w},b,\xi)$ from the optimization described as
\begin{align}
\centering
\min_{\textbf{w},b,\xi}~~ &\frac{1}{2} \left \| \mathbf{w} \right \|^2 +C\sum \xi_i.\label{eq:optimization_svm}\\
\textup{subject to}~~~&y_i(\mathbf{w}\cdot\Phi(\mathbf{x_i}+b)+\xi_i \geq 1 ~\textup{and}~ \xi_i \geq 0 \notag\\
&for~i = 1,...,N.\notag
\end{align}

Notice that $C$ is a variable that compensates for the size of $\mathbf{w}$  and the sum of $\xi$ in order to avoid over fitting. For numerical computation purposes, the dual form of (\ref{eq:optimization_svm}) is used. This dual form is defined in (\ref{eq:optimization_svm_dual})-(\ref{eq:hyperplane_svm_dual}) as
\begin{align}
\min_{\alpha}~~ &\frac{1}{2} \sum_{i,j=1}^{N} y_i y_j \alpha_i \alpha_j K(\mathbf{x_i},\mathbf{x_j}) - \sum_{i=1}^{N} \alpha_i,\nonumber\\
\textup{subject to}~& \sum_{i=1}^{N} y_i \alpha_i = 0~\textup{and}~ 0\leq \alpha_i \leq C, ~\nonumber\\  &for~i=1,..,N\label{eq:optimization_svm_dual}
\end{align}
where
\begin{align}
K(\mathbf{x_i},\mathbf{x_j}) &= \mathbf{\Phi(x_i)}\cdot\mathbf{\Phi(x_j)}\label{eq:kernel_svm_dual}\\
\mathbf{w} &= \sum_{i=1}^{N} \alpha_i \mathbf{\Phi(x_i)}\label{eq:w_svm_dual}\\
f(\mathbf{x}) &= \sum_{i=1}^{N} y_i\alpha_i K(\mathbf{x_i},\mathbf{x_j}) + b.\label{eq:hyperplane_svm_dual}
\end{align}

Eq.~(\ref{eq:hyperplane_svm_dual}) is used as the decision function, defining which label should be applied to a specific test sample. If $f(\mathbf{x}) \leq 0$ the prediction is labeled +1, otherwise the prediction is labeled 0. In this case, the binary choice of labels can be chosen to be anything that ones need to classify.  Because this paper deals with several labels, an approach of ``one-vs-all", derived from the standard SVM, will be shown in section \ref{sec:alg}.

\section{Problem Formulation}\label{sec:PF}

In the context of this paper, we consider a 1-dimensional road, having a length denoted as $D\in \mathbb{R}_{>0}$. Let $\mathcal{S}=\{S_1,\cdots,S_{N_S}\}$ be a set of $N_S\in \mathbb{Z}_{>0}$ sensors placed along the road at the locations $\mathcal{R}=\{R_1,\cdots,R_{N_S}\}$, where each element in $\mathcal{R}$ is unique and valued in the range $(0,D)$.

As a target passes the sensor, the sensor collects the target's information. This information includes the velocity of the target (denoted as $v$) and a timestamp representing when the target passes the sensor (denoted as $t$). The information collected about the target is disassociated with the target, meaning that the target for which the information was received cannot be directly identified using the information. 

Let $\mathcal{A}=\{A_1,\cdots,A_{N_A}\}$ represent a set of $N_A\in \mathbb{Z}_{>0}$ targets. It is assumed that each target passes each sensor exactly one time. The sensors store their collected information in the matrices $V\in \mathbb{R}^{N_{A}\times N_{S}}$ and $T\in \mathbb{R}^{N_{A}\times N_{S}}$. The $m^{th}$ velocity and timestamp measurements obtained by the $n^{th}$ sensor is recorded as $V_{n,m}\in V$ and $T_{n,m}\in T$, respectively (where $n\in \{1,\cdots,N_S\}$ and $m\in \{1,\cdots,N_A\}$). Let the information $\mathcal{X}\subset \mathcal{R}^1$ be the collection of velocity data $V$ and timestamp information $T$, organized such that each element $X_{n}^m\in \mathcal{X}$ is a set of data conaining the values $V_{n,m}$ and $T_{n,m}$. More precisely, the set of all observations from the sensor network is represented as $\mathcal{X} = \{X_1^1, X_1^2,...,X_1^{N_A},...,X_{N_S}^{N_A}\}$.


\begin{figure}[ht!]
\centering
\captionsetup{justification=centering}
\includegraphics[width=8cm]{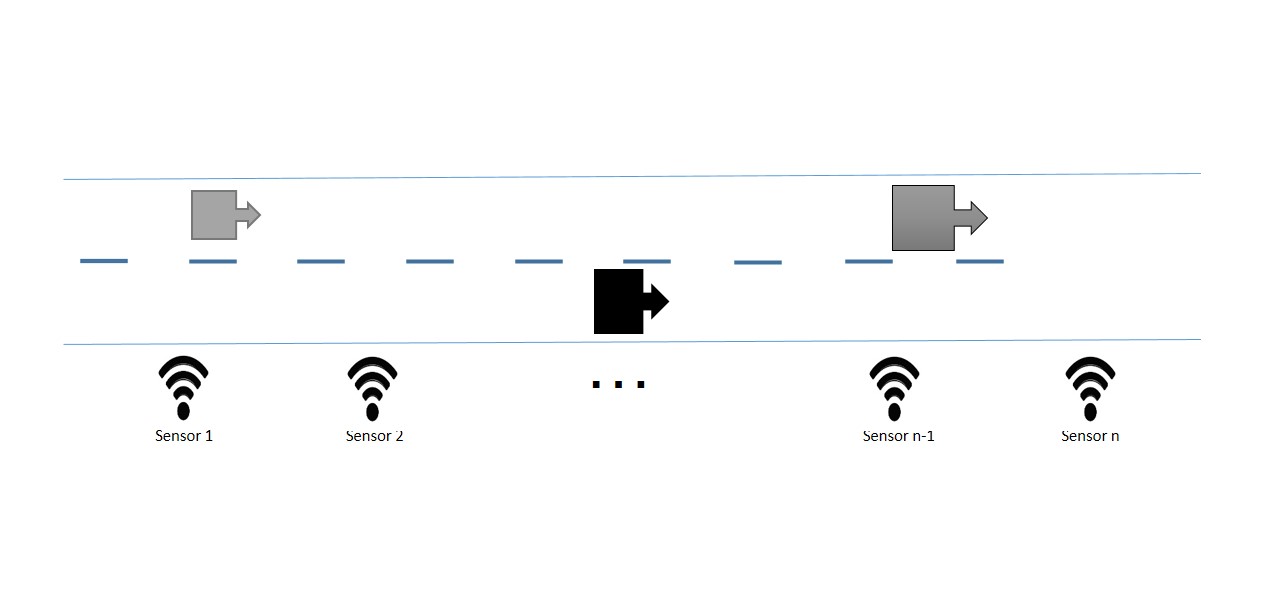}
\caption{Sensor distribution and road configuration}
\label{fig:roadNW}
\end{figure}

Let $\mathcal{X}_{ideal} \subset \mathcal{R}^2$ be the desired outcome of the form given by
\begin{align}\label{eq:ideal}
\mathcal{X}_{ideal}=
\begin{Bmatrix}
X_1^1&\cdots&X_{N_S}^1\\
\vdots&\ddots&\vdots\\
X_1^{N_A}&\cdots&X_{N_S}^{N_A}\\
\end{Bmatrix},
\end{align}
where each row of $\mathcal{X}_{ideal}$ represents the measurements of all sensors regarding the same target. For example, the $i$th row of $\mathcal{X}_{ideal}$ is the dataset associated with target $A_i$ observed by different sensors at different time instances. Because the targets' velocities fluctuate as they move along the road network, the  sequence of targets observed by one sensor may not be the same as that observed by another one, leading a necessary data pattern recognition problem. More precisely, given a set of observations from all sensors as $\mathcal{X} = \{X_1^1, X_1^2,...,X_1^{N_A},...,X_{N_S}^{N_A}\}$, \footnote{In this case, $X_n^m$ means the $m^{th}$ sample collected by the $n^{th}$ sensor, ($m$ does not identify a specific target).} our goal is to classify the data $\mathcal{X}$ in the desired form of $\mathcal{X}_{ideal}$ as in~\eqref{eq:ideal}.
For the simplicity of representation, we assume that no false alarm or missing detection will occur although the proposed methods can be augmented appropriately to deal with false alarm and missing detection.

To evaluate the performance of the proposed method, labeled data is needed to obtain the percentage of true association between targets and their measurements. Let the information $\mathcal{Y}$ be the collection of information that includes the true association of targets to their measurements. The information $\mathcal{Y}$ is structured similarly to $\mathcal{X}$ except that each element $Y_n^m \in \mathcal{Y}$ also includes the value $A_i$, which describes the target identification that the measurement corresponds to.

Algorithm \ref{alg:data-gen} is the pseudo code that describes how multiple datasets of $\mathcal{Y}$ are generated. The data sets of $\mathcal{Y}$ are generated in iterations. Let $N_K$ be the total number of iterations and let $K$ be the current iteration step. In each iteration step, there is a sub-iterative process. The sub-iterative process generates and records the trajectory for a single target. The trajectory is generated by first defining the initial time and velocity values for the target (line 5 and 6 of Algorithm 1). These values are selected from a uniform random variable, where $(v_{min}, v_{max})$ and $(t_{min},t_{max})$ indicate, respectively, the upper and lower bound for initial velocity and time. A unique trajectory is generated for the profile $V_i(t)$ by the function $g(\cdot,\cdot)$, which is not fully described here (line 7 of Algorithm 1). Then, according to this trajectory the function $h(\cdot,\cdot,\cdot)$ is used to insert measurements into the appropriate elements of $\mathcal{Y}_K$ (line 8 of Algorithm 1). The velocity and time measurements associated with target $A_i$ are inserted into the element location of $\mathcal{Y}_K$ according to the sensor positions $\mathcal{R}$. After inserting all measurements, the information $\mathcal{Y}_K$ is saved (line 10 of Algorithm 1). 

\begin{algorithm}
\caption{Data generation}\label{alg:data-gen}
\begin{algorithmic}[1]
\State{Output: Generate multiple $\mathcal{Y}$ from $N_S$ sensors on a road with length $D$}
\State $\mathcal{R} =  \{R_1,R_2,...,R_m\}$;
\For {$K =1:N_K$}  		 
\For {$i =  1:N_A$} 		
\State  $v_{o_i}= \sim U [(v_{min},v_{max})]$;
\State  $t_{o_i} = \sim U [(t_{min},t_{max})]$;

\State $V_i(t) = g(v_{o_i},t_{o_i})$;
\State $\mathcal{Y}_K\leftarrow h(V_i(t),A_i,\mathcal{R})$;
\EndFor 
\State Save $\mathcal{Y}_K$;
\EndFor
\end{algorithmic}
\end{algorithm}

\section{Algorithm}\label{sec:alg}
To solve the aforementioned problem, the objective of this section is to derive a set of machine learning algorithms that classify the information $\mathcal{X}$, such that each target-measurement pair can be identified. In addition, we are interested in conducting a comparison of the performance of these algorithms for different datasets. Using the appropriate machine learning techniques, it is expected that the random data $\mathcal{X}$ can be re-organized in order to achieve the mapping described as
\begin{align}
\Theta:(\mathcal{X} \subset \mathcal{R}^1) \mapsto (\mathcal{X}_{ideal} \subset \mathcal{R}^2),\label{eq:mapping}
\end{align}
where each row of $\mathcal{X}_{ideal}$ represents a set of all measurements associated to the $m^{th}$ target. The performance of the proposed machine learning algorithms will be evaluated by the associated accuracy levels. 


\subsection{Data Pre-processing}
\begin{figure}[h!]
\centering
\captionsetup{justification=centering}
\includegraphics[width=8cm]{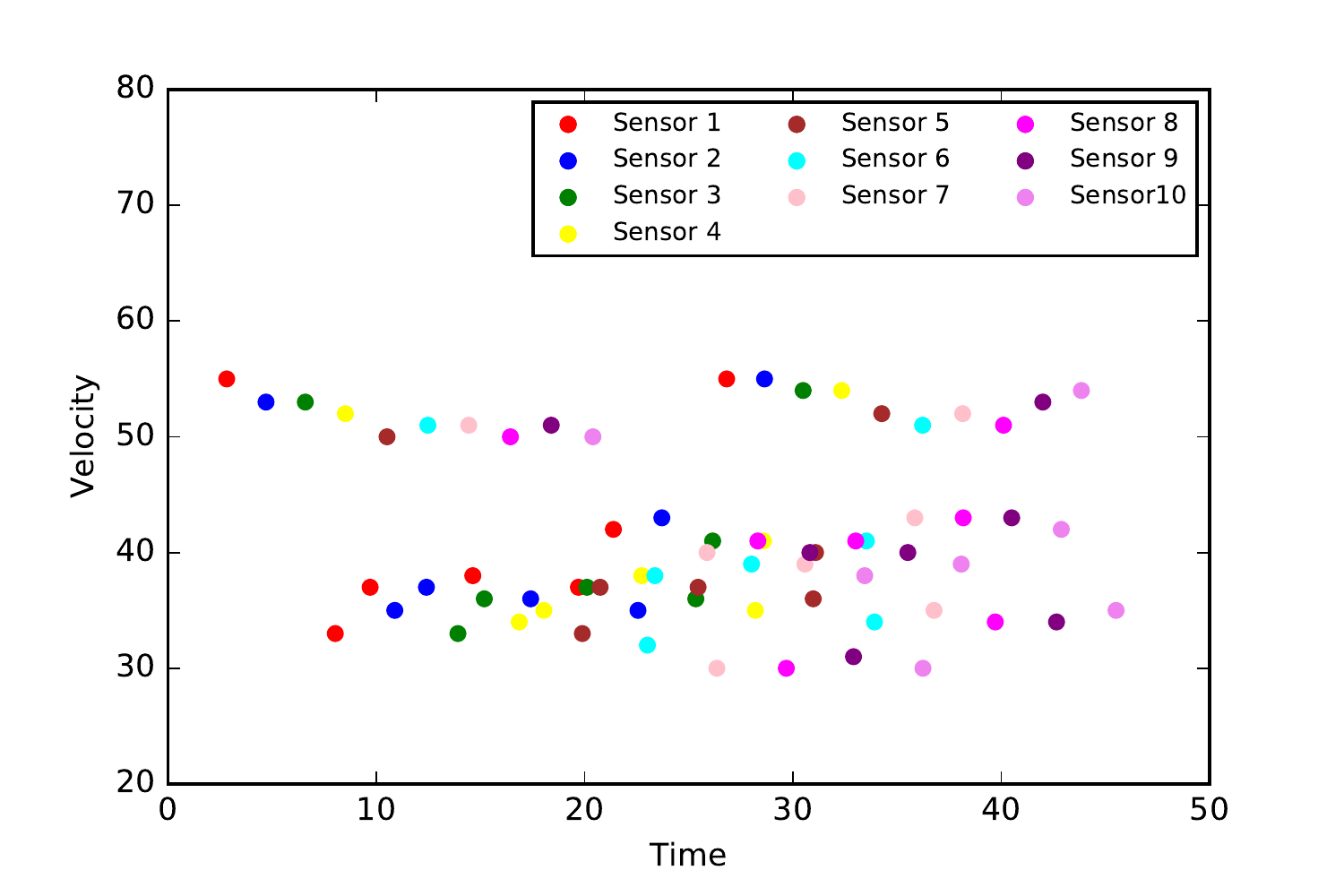}
\caption{Dataset with a large velocity variance}
\label{fig:firstDataset}
\end{figure}

\begin{figure}[h!]
\centering
\captionsetup{justification=centering}
\includegraphics[width=8cm]{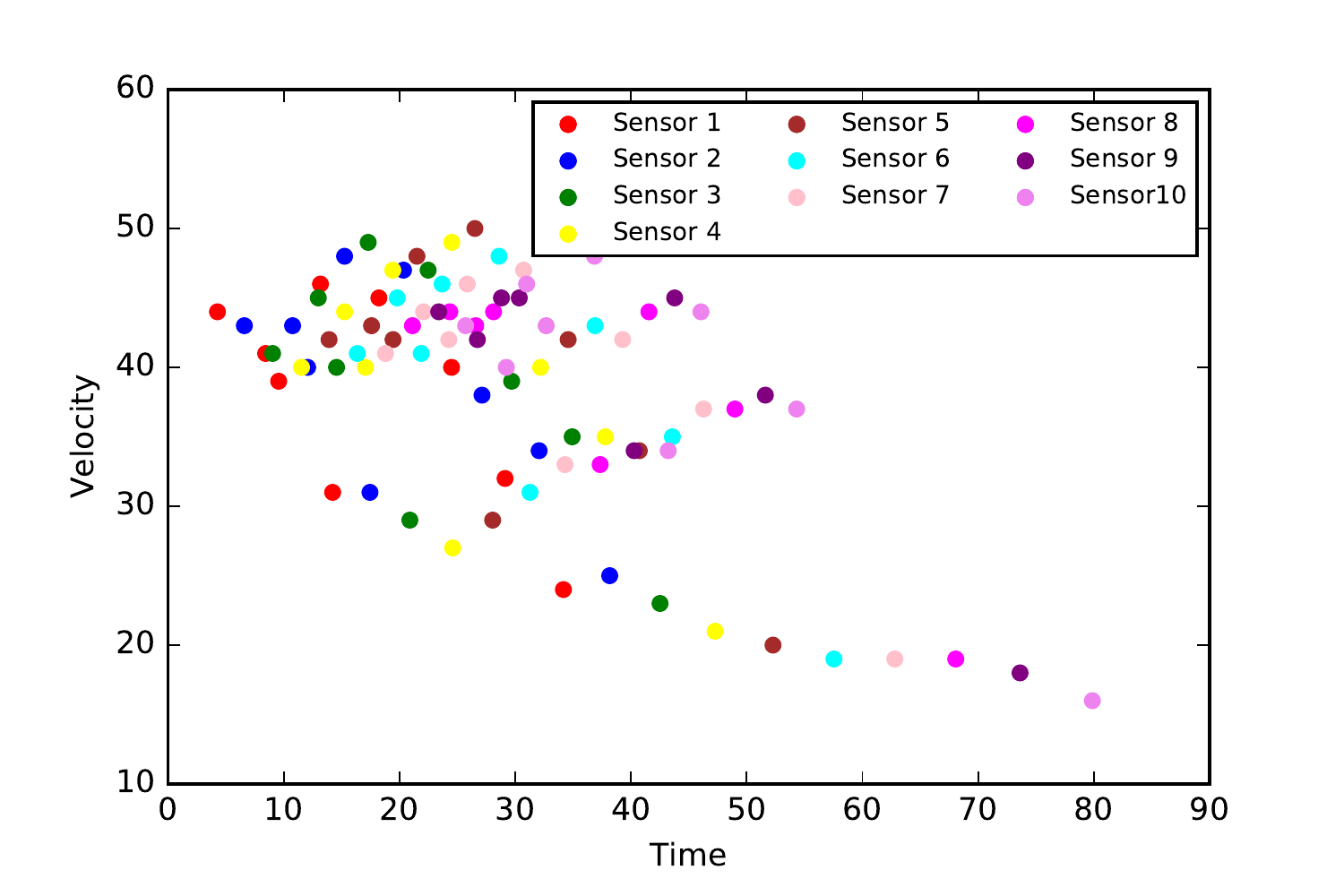}
\caption{Dataset with a medium velocity variance}
\label{fig:secondDataset}
\end{figure}

\begin{figure}[h!]
\centering
\captionsetup{justification=centering}
\includegraphics[width=8cm]{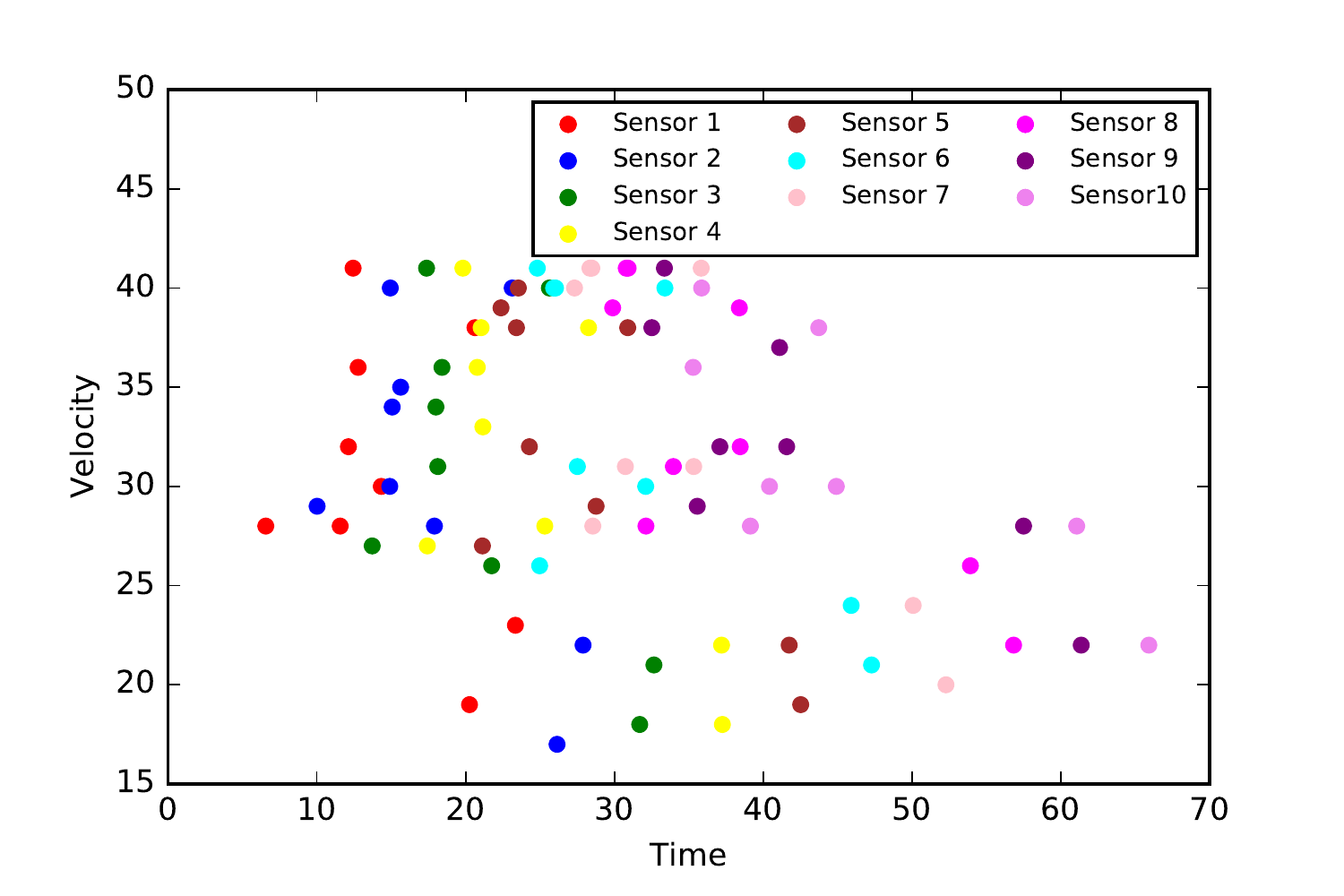}
\caption{Dataset with a small velocity variance}
\label{fig:thirdDataset}
\end{figure}

Figures \ref{fig:firstDataset}, \ref{fig:secondDataset},~\ref{fig:thirdDataset} show the case where overlapping between measurements in $\mathcal{X}$ occur without any preprocessing of the data for different velocity variances. In fact, the overlapping of dataset becomes more significant if more targets are involved. By using the raw data $\mathcal{X}$, extracting data patterns using machine learning algorithms, such as K-means and SVM, is difficult. Moreover, if there exists overlapping of data points from different targets, mis-classification of these points into the same category is well expected. To obtain more accurate data patterns, an appropriate pre-processing of $\mathcal{X}$ is essential.

The method for pre-processing the data is to project the  measurements in $\mathcal{X}$ and estimate what would be that target's measurement obtained by sensor $S_1$. A notation for such a mapping is given by
\begin{align}
&\mathcal{F}:(S_i \in \mathcal{S}) \mapsto (\mathcal{S}_1 \in \mathcal{S})\\
\forall~ (0&\leq n\leq N_S) \textup{ and } \forall~ (0\leq m\leq N_A).\nonumber \label{eq:backmapping}
\end{align}
The data obtained after pre-processing is represented as
\begin{equation} \label{eq:afterPreprocessing}
\mathcal{X'} = \begin{Bmatrix}
{X'}_1^1&\cdots&{X'}_{N_S}^1\\
\vdots&\ddots&\vdots\\
{X'}_1^{N_A}&\cdots&{X'}_{N_S}^{N_A}\\
\end{Bmatrix}.
\end{equation}

In this approach the sensor $S_1$ (the first sensor) is taken as a reference. All other sensor readings are used to generate an estimate for what the reading at sensor $S_1$ would be. For each sensor $S_{n}$, we have a velocity ($V_{n,m}$), time ($T_{n,m}$) and position ($R_n$). We use the velocity and position of the sensor to calculate a time value that represents when the target had passed sensor $S_1$. The new information is represented as the set of data ${X'}_n^m$ consisting of velocity $(V_{n,m})$ and time (${T'}_{n,m}$). 


\begin{figure}[h!]
\centering
\captionsetup{justification=centering}
\includegraphics[width=8cm]{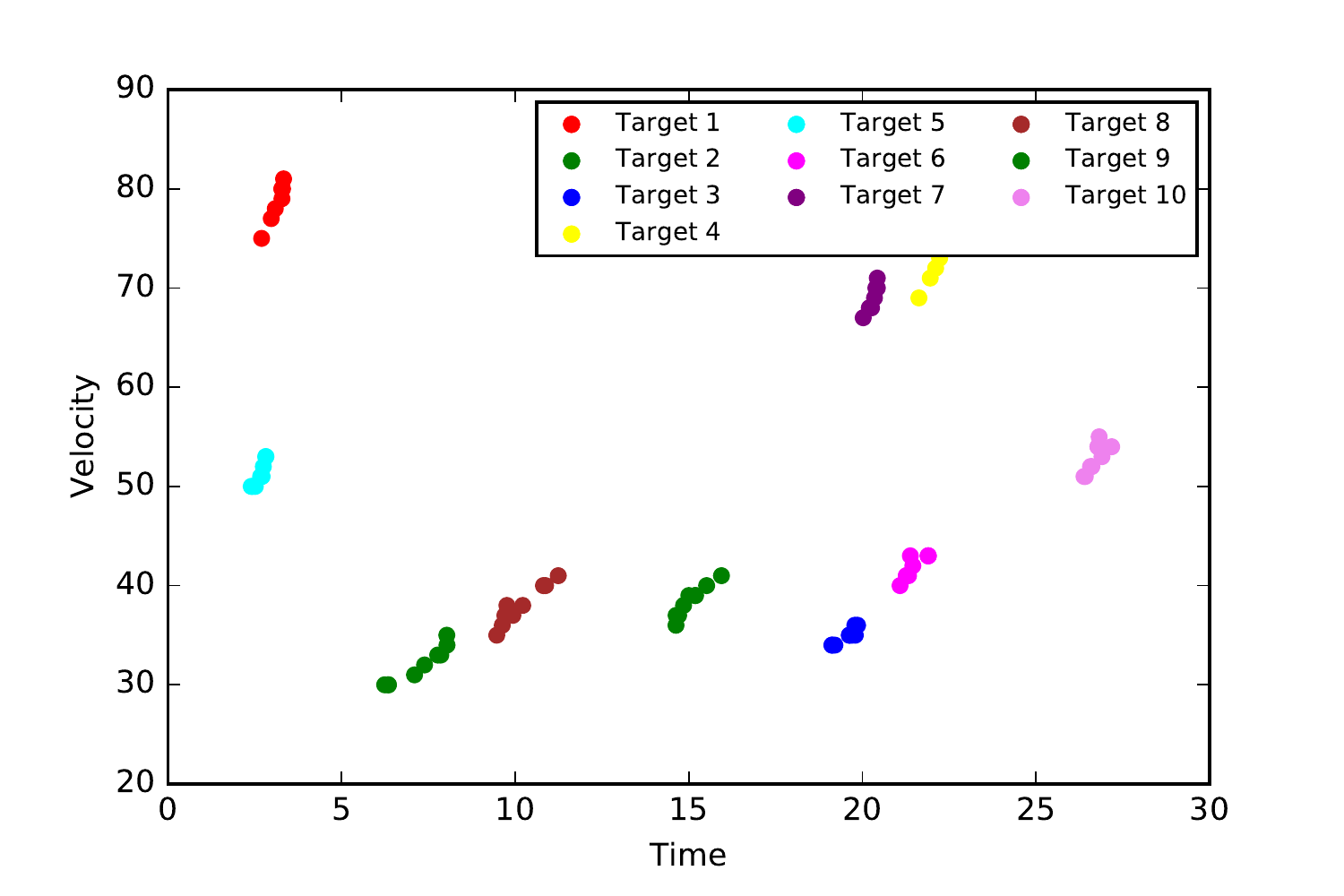}
\caption{Dataset with a large velocity variance after pre-processing}
\label{fig:ProcessedDataset1}
\end{figure}

\begin{figure}[h!]
\centering
\captionsetup{justification=centering}
\includegraphics[width=8cm]{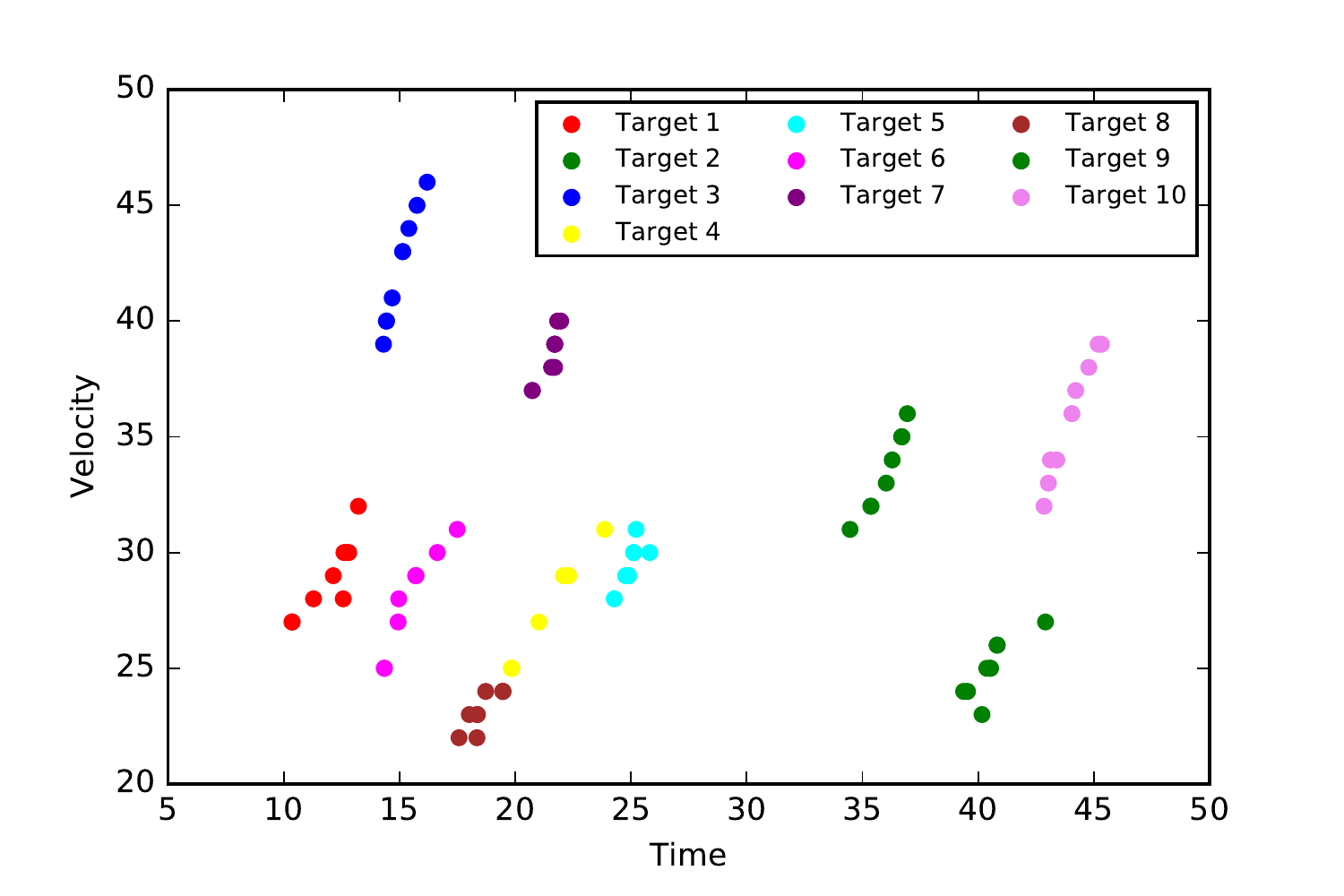}
\caption{Dataset with a medium velocity variance after pre-processing}
\label{fig:ProcessedDataset2}
\end{figure}

\begin{figure}[h!]
\centering
\captionsetup{justification=centering}
\includegraphics[width=8cm]{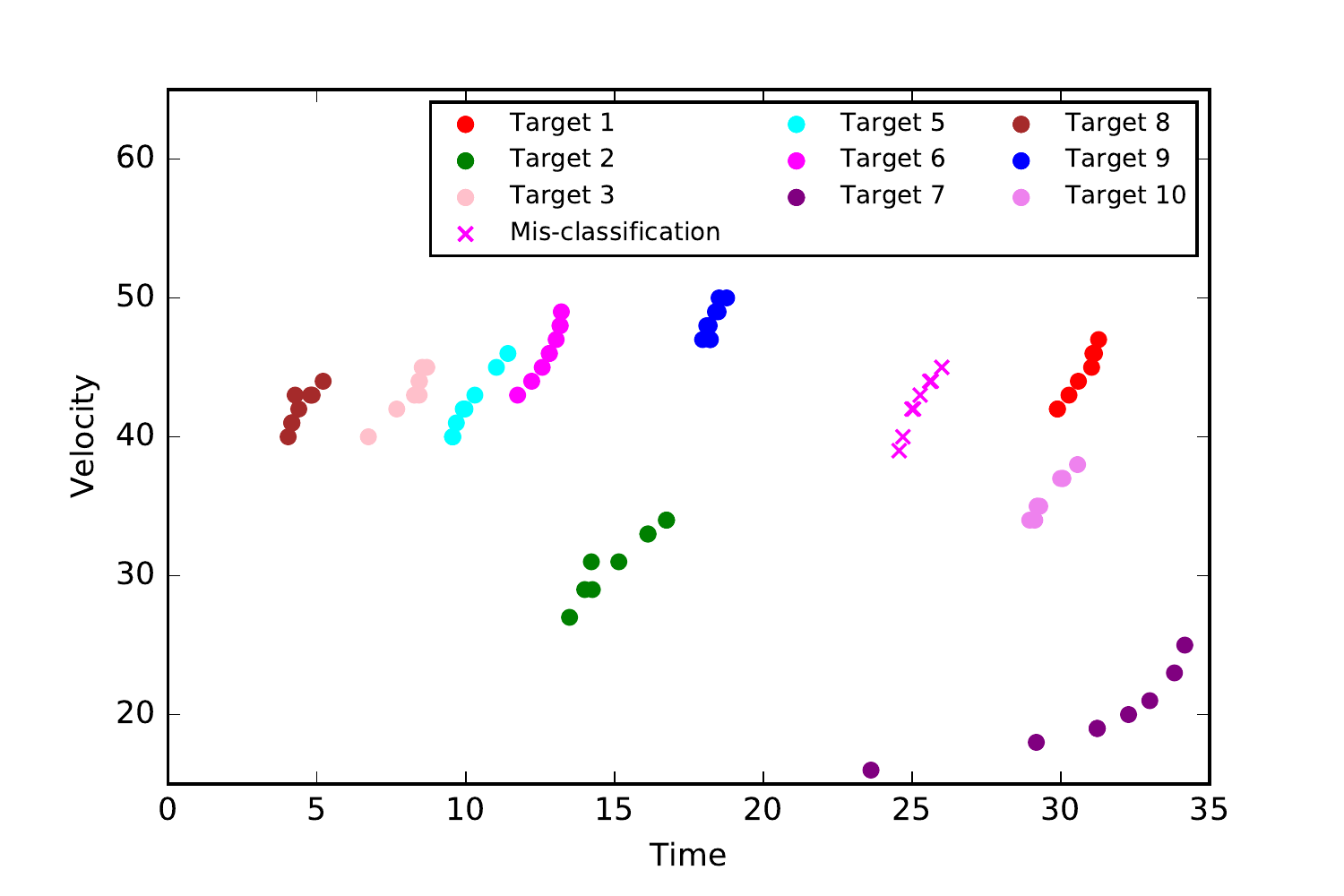}
\caption{Dataset with a small velocity variance after pre-processing}
\label{fig:ProcessedDataset3}
\end{figure}

The definition of this mapping is given by
\begin{align} \label{eq:pre-processing} 
{T'}_{n,m} =T_{n,m} ~-~ \frac{(R_n-R_1)}{V_{n,m}}.
\end{align}
Eq.~\ref{eq:pre-processing} describes the backward extrapolation of the time from the $n^{th}$ sensor to sensor $S_1$. For example, the 4th measurement obtained by sensor $S_2$ includes $V_{2,4}$ = 20 m/s and $T_{2,4}$ = 50 s. Let the distance between sensor $S_1$ and $S_2$ (evaluated as $R_2 - R_1$) be 100 units. The estimated time associated with the measurement $X_2^4$ is then calculated as ${T'}_{n,m}$ = 45 s. Conducting the same pre-processing strategy to the information $\mathcal{X}$ results in the converted time estimation of all measurements with respect to sensor $S_1$. Figures \ref{fig:ProcessedDataset1}, \ref{fig:ProcessedDataset2}, \ref{fig:ProcessedDataset3} show the plot of the datasets used in Figures \ref{fig:firstDataset}, \ref{fig:secondDataset}, and~\ref{fig:thirdDataset} respectively after applying the proposed data pre-processing method. As can be observed from Figures \ref{fig:ProcessedDataset1}, \ref{fig:ProcessedDataset2}, \ref{fig:ProcessedDataset3} that the processed data yields some patterns that can be potentially recognized by machine learning algorithms, such as K-means and SVM.


\subsection{Classification Algorithms}
Two different algorithms are used for data classification. A K-means clustering algorithm is used as the primary approach because the described problem fits the classic example of an unsupervised learning problem. When an unlabeled data is clustered with K means, it identifies the clusters and assigns each of the data points to one of the clusters. Each cluster here represents a target. A slightly modified version of K-means algorithm, called K-means++ algorithm \cite{arthur2007k}, is used to provide better performance as shown in the simulation examples in Section~\ref{sec:simu}. For the K-means++ algorithm, a center $k_1$ is chosen randomly as the first step. Then the next cluster center $k_{j+1}$ is chosen, with $ X_n^m \in \mathcal{X}'$, according to the probability given by, 
\begin{equation} \label{eq:Kmns++Probability}
\frac{D_s(X_n^m)^2}{\sum_{X_n^m \in X}D_s(X_n^m)^2}
\end{equation}
where $D_s(X_n^m)$ denotes the shortest distance from a data point $X_n^m$ to the closest center $k_{j-1}$ already chosen.

An alternative algorithm chosen to solve this problem is SVM. The SVM algorithm discussed in section \ref{sec:SVM} is an ideal classifier which involves classification between two categories. In order to develop a multi-class SVM classifier a ``one-vs-all' classifier is used. There are $N_S$ different binary classifiers built such that for any target $A_i$, all the points in class $A_i$ are fed as positive samples, and all points not in $A_i$ are fed as negative samples. For targets $A_i$ with classifier functions $f_i$, the output classifier function $f(x)$ for a new data $x$ can be obtained via the following operation:
\begin{equation}\label{eq:SVM-1}
f(x) = arg~ \max_i~f_i(x).
\end{equation}

\subsection{Recovering the original dataset}

In the previous section, the measurements were classified for each target based on the pre-processed data. Predicting and classifying the pre-processed data is an intermediate solution to the problem and does not directly give any meaning to our original data classification. When the dataset is pre-processed and projected backwards with reference to sensor $S_1$, the result obtained is given by \eqref{eq:afterPreprocessing}, where ${X'}_n^m$ consists of information $V_{n,m}$, ${t'}_{n,m}$ and corresponds to sensor $S_n$ located at $R_n$. Therefore, the original time $t_i^j$ can be obtained via the following operation:
\begin{equation*}
t_{n,m} = {t'}_{n,m} + \frac{R_n-R_1}{V_{n,m}}.
\end{equation*}
With the original time $t_{n,m}$, velocity $V_{n,m}$ and position of the sensor $R_n$, the final classified set $\mathcal{X}''$ can be obtained. Each unique row in $\mathcal{X}''$ consists of measurements associated to a single target. In particular, $\mathcal{X}''$ can be written in the desired form given in~\eqref{eq:ideal}. Algorithm~\ref{alg:K-means} is the pseudo code that describes the proposed learning algorithm that includes the data preprocessing, classification, and recovering as described in the previous part of the section.

\begin{algorithm}
\caption{Learning algorithm}\label{alg:K-means}
\begin{algorithmic}[1]
\State{Input: $ \mathcal{X} $}
\State{Output: Obtain ~$\mathcal{X}'' = \begin{bmatrix}
 {X}_1^1 &{X}_2^1&\ldots&{X}_m^1\\
 \vdots&\vdots&\vdots&\vdots\\
 {X}_1^n &{X}_2^n&\ldots&{X}_m^n\\
 \end{bmatrix}$}
\While{not at the end of the file} 
\State $X[m \times n] \leftarrow \begin{bmatrix}
V_1^1 &t_1^1  &P_1 \\ 
 \vdots& \vdots &\vdots \\ 
 V_1^n&t_1^n  & P_m
\end{bmatrix}$;
\EndWhile
\While {not at the end of X[$m\times n$]}
\State $tp_i^j = t_i^j - \frac{P_i- P_1}{V_i^j};$
\State $\mathcal{X'}$ $\leftarrow V_i^j, {t'}_i^j, P_i;$
\EndWhile
\State Create a model for K-means++ / one-versus-all SVM
\State Run / Train the model
\While {not at the end of $\mathcal{X'}$}
\State $t_i^j = {t'}_i^j + \frac{P_i-P_1}{V_i^j};$
\State $\mathcal{X}'' \leftarrow V_i^j, {t}_i^j, P_i;$
\EndWhile
\State return $\mathcal{X}''$
\end{algorithmic}
\end{algorithm}

\section{Simulation}\label{sec:simu}

In this section, we provide some simulation examples to demonstrate the effectiveness of the proposed machine learning algorithms and compare their performances. To show the benefits of pre-processing, the proposed algorithms are applied to both the raw data $\mathcal{X}$ and the preprocessed data $\mathcal{X}'$. In addition, the accuracy levels of these algorithms are formally compared. To fully investigate the performance of the proposed algorithms, we select a variety of target motions, i.e., the velocity profile of targets falls into different ranges. The parameters used in the simulation examples are given in Tables~\ref{tb:dataset_l},~\ref{tb:dataset_m}, and~\ref{tb:dataset_s}. Figures \ref{fig:firstDataset},~\ref{fig:secondDataset} and \ref{fig:thirdDataset} show the plot of the dataset obtained via the simulation parameters in Tables \ref{tb:dataset_l}, \ref{tb:dataset_m} and \ref{tb:dataset_s}, respectively.

\begin{table}[!h]
\begin{center}
\caption{Parameters with a large velocity variance}
\label{T:dataset3}
\begin{adjustbox}{width= 3.5in,totalheight=0.3in}
\begin{tabular}{| c | c | c | c | c |}
\hline
\textbf{$N_A$} & \multicolumn{1}{ c |}{D}& \multicolumn{1}{ c |}{$S_i$}& \multicolumn{1}{ c |}{\textbf{ $V_i$}}& \multicolumn{1}{ c |}{\textbf{$T_i$ }} \\ \hline
$10$ & $1000$& every $100(i-1)$ & $[30, 80] $& $ [1, 30] s $\\ \hline
\end{tabular}\label{tb:dataset_l}
\end{adjustbox}
\end{center}
\end{table}

\begin{table}[!h]
\begin{center}
\caption{Parameters with a medium velocity variance}
\label{T:dataset3}
\begin{adjustbox}{width= 3.5in,totalheight=0.3in}
\begin{tabular}{| c | c | c | c | c |}
\hline
\textbf{$N_A$} & \multicolumn{1}{ c |}{D}& \multicolumn{1}{ c |}{$S_i$}& \multicolumn{1}{ c |}{\textbf{ $V_i$}}& \multicolumn{1}{ c |}{\textbf{$T_i$ }} \\ \hline
$10$ & $1000$& every $100(i-1)$ & $[20, 50] $& $ [1, 30] s $\\ \hline
\end{tabular}\label{tb:dataset_m}
\end{adjustbox}
\end{center}
\end{table}

\begin{table}[!h]
\begin{center}
\caption{Parameters with a small velocity variance}
\begin{adjustbox}{width= 3.5in,totalheight=0.3in}
\begin{tabular}{| c | c | c | c | c |}
\hline
\textbf{$N_A$} & \multicolumn{1}{ c |}{D}& \multicolumn{1}{ c |}{$S_i$}& \multicolumn{1}{ c |}{\textbf{ $V_i$}}& \multicolumn{1}{ c |}{\textbf{$T_i$ }} \\ \hline
$10$ & $1000$& every $100(i-1)$ & $[20, 40] $& $ [1, 20] s $\\ \hline
\end{tabular}\label{tb:dataset_s}
\end{adjustbox}
\end{center}
\end{table}



\begin{figure}[h!]
\centering
\captionsetup{justification=centering}
\label{fig:kmeansUnprocessed}
\includegraphics[width=8cm]{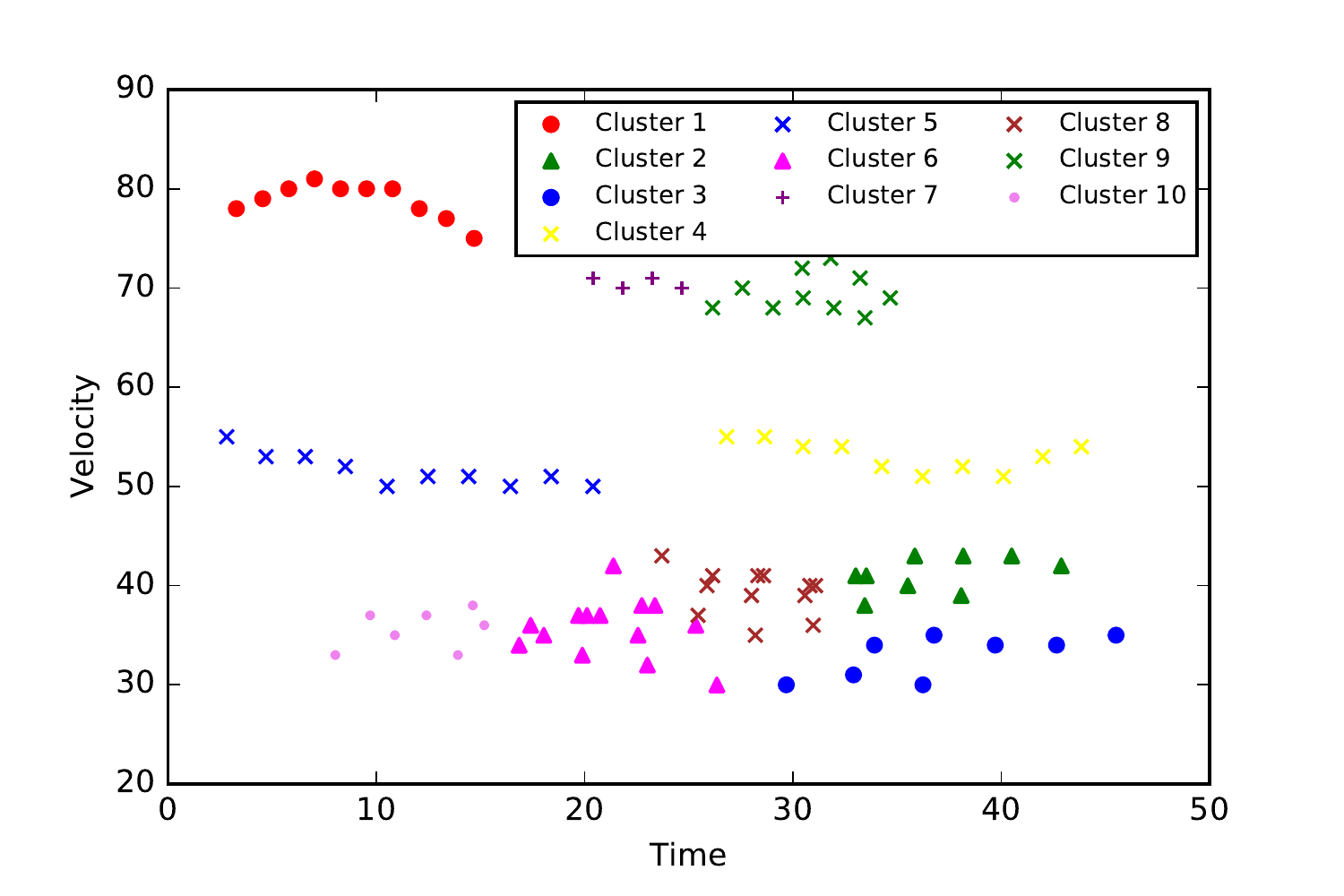}
\caption{K-means++ classification of unprocessed data in Fig.~\ref{fig:firstDataset}.}
\label{fig:KmeansUnprocessedData}
\end{figure}

Figure \ref{fig:KmeansUnprocessedData} shows the classification of unprocessed data generated using the parameters in Table \ref{tb:dataset_l}. The classification of the unprocessed data is obtained using the K-means++ clustering algorithm. It can be observed from Figures~\ref{fig:firstDataset} and \ref{fig:KmeansUnprocessedData} that the accuracy is very low because of the significant overlapping of raw data in the absence of data pre-processing. For the data generated using the parameters shown in Tables \ref{tb:dataset_m} and \ref{tb:dataset_s}, same worse performance is obtained using K-means++ clustering algorithm. Therefore, pre-processing of the raw datasets is necessary to obtain more accurate data patterns. 
\begin{figure}[h!]
\centering
\captionsetup{justification=centering}
\includegraphics[width=8cm]{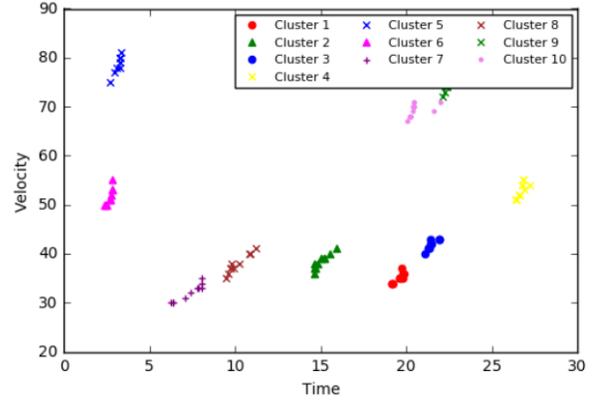}
\caption{K-means++ classification of pre-processed data in Fig.~\ref{fig:ProcessedDataset1}.}
\label{fig:kmeansProcessed1}
\end{figure}

\begin{figure}[h!]
\centering
\captionsetup{justification=centering}
\includegraphics[width=8cm]{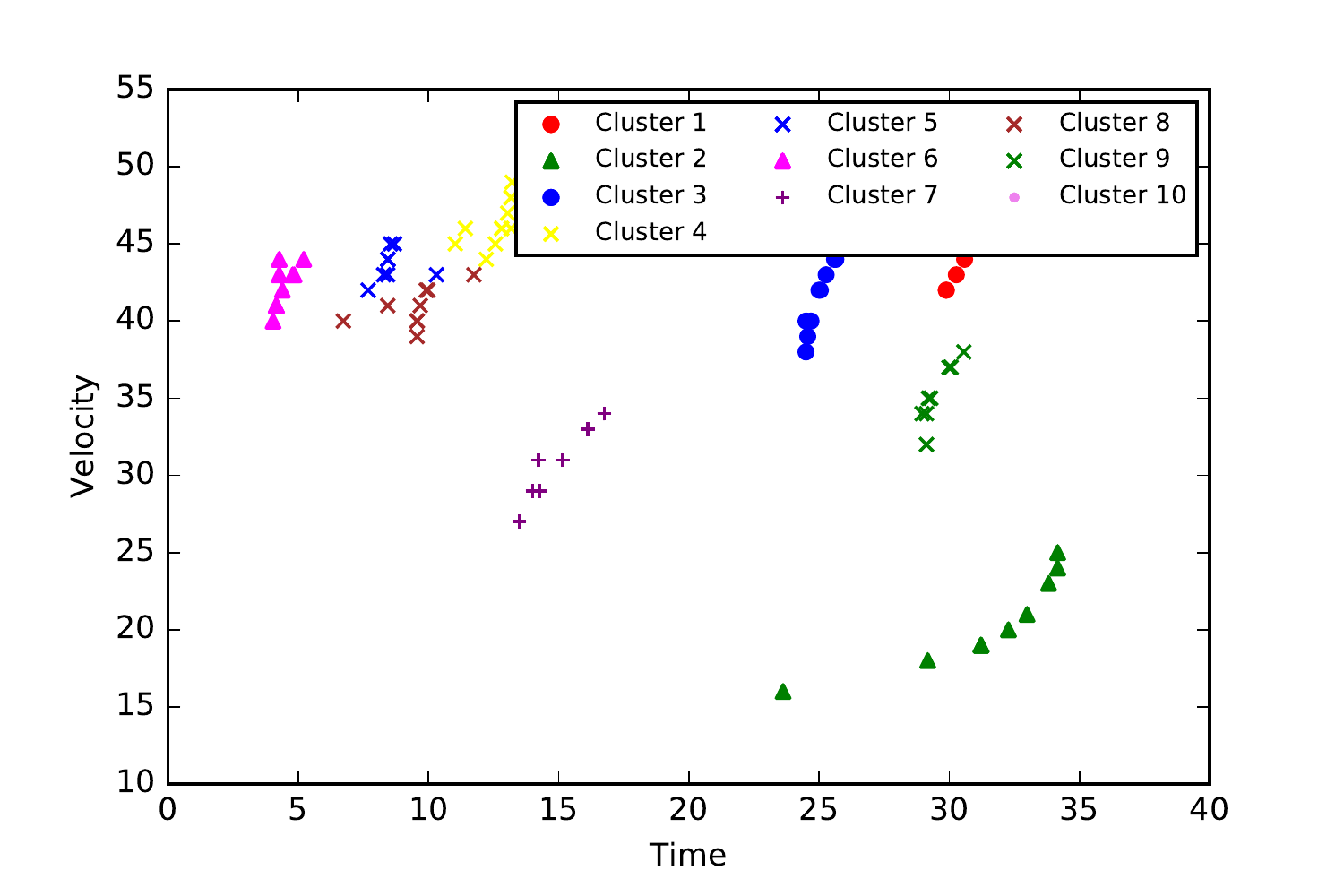}
\caption{K-means++ classification of pre-processed data in Fig.~\ref{fig:ProcessedDataset2}.}
\label{fig:kmeansProcessed3}
\end{figure}

\begin{figure}[h!]
\centering
\captionsetup{justification=centering}
\includegraphics[width=8cm]{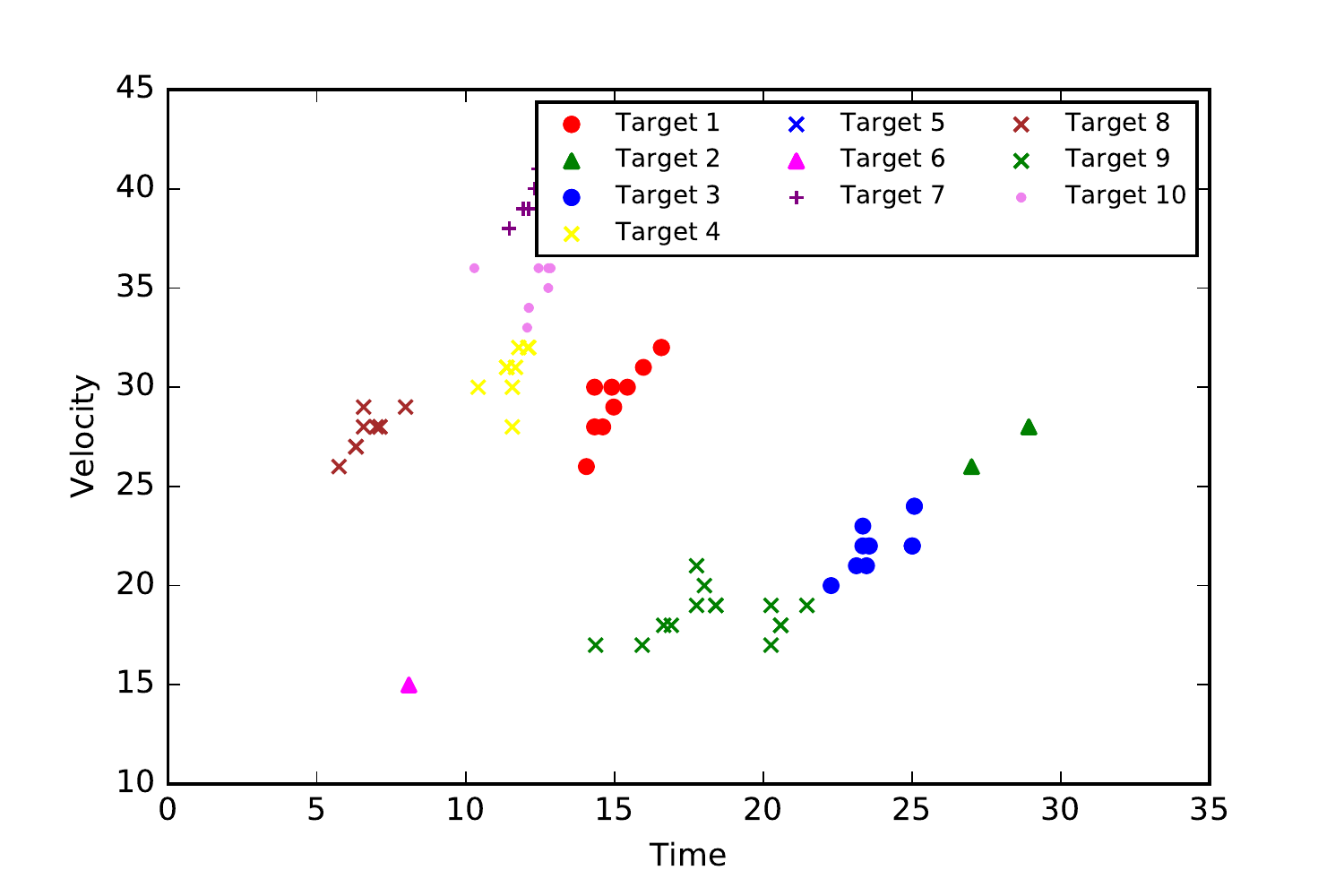}
\caption{K-means++ classification of pre-processed data in Fig.~\ref{fig:ProcessedDataset3}.}
\label{fig:kmeansProcessed4}
\end{figure}

Figures \ref{fig:kmeansProcessed1}, \ref{fig:kmeansProcessed3} and \ref{fig:kmeansProcessed4} shows the result of the classification obtained from the K-means++ clustering algorithm on the pre-processed datasets, generated using the parameters shown in Tables \ref{tb:dataset_l}, \ref{tb:dataset_m} and \ref{tb:dataset_s}, respectively. It can be observed from Figures \ref{fig:kmeansProcessed1} and \ref{fig:kmeansProcessed3} that the accuracy of the classification obtained from K-means++ clustering algorithm is very high. Most data points are classified accurately with very few mis-classifications. This high accuracy can be attributed to the wide velocity and time distribution of the two datasets as shown in Figures \ref{fig:ProcessedDataset1} and \ref{fig:ProcessedDataset3}. The accuracy of the K-means++ algorithm on the dataset obtained using Table \ref{tb:dataset_s} is poor and is shown in Figure 10. The poor accuracy is due to the narrow distribution of velocity and time in the dataset. This results in more overlapping of data points even after the pre-processing is performed on the dataset. 


\begin{figure}[h!]
\centering
\captionsetup{justification=centering}
\includegraphics[width=8cm]{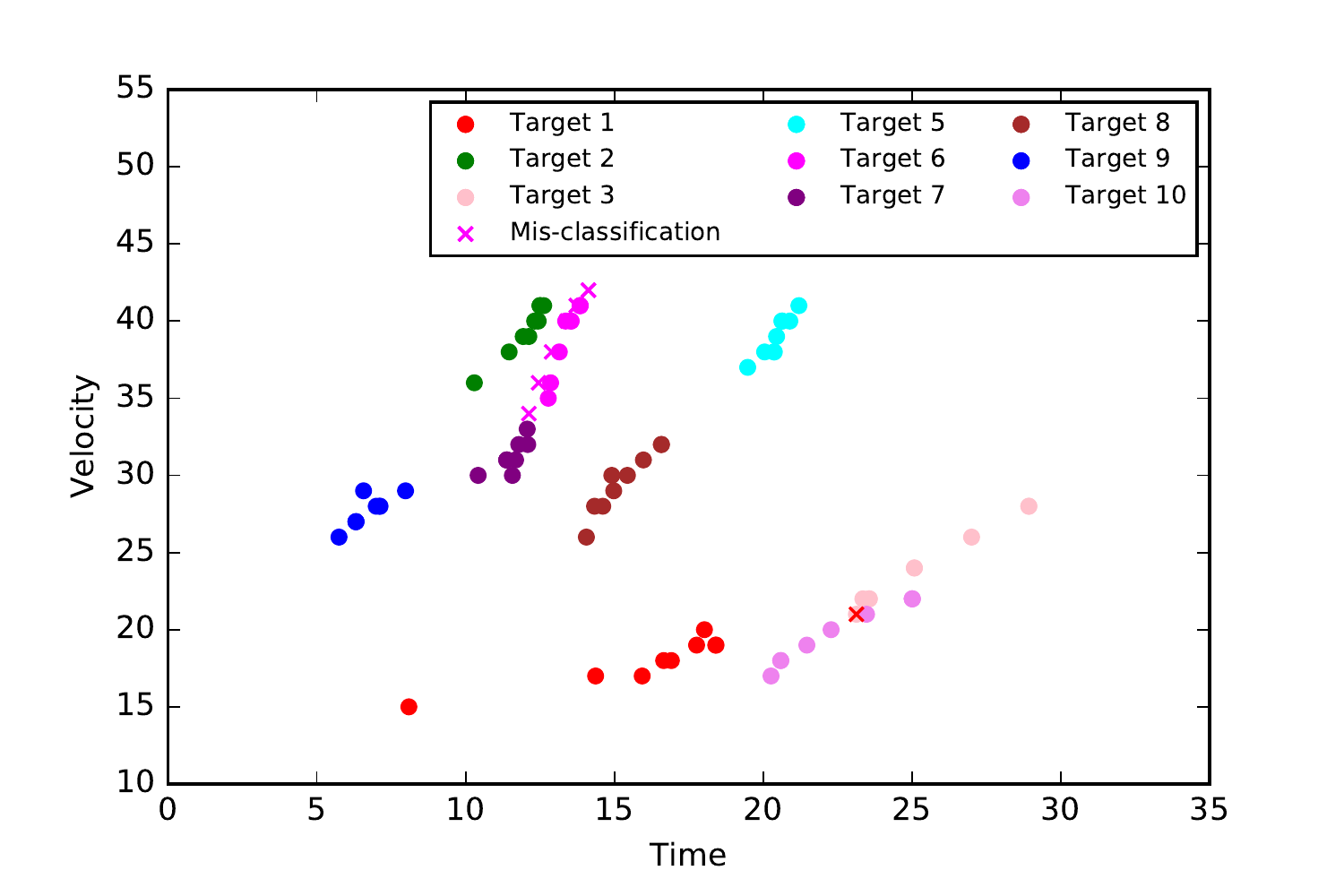}
\caption{Quadratic SVM (QSVM) classification of pre-processed data in Fig.~\ref{fig:ProcessedDataset3}.}
\label{fig:SVMProcessed4}
\end{figure}

To overcome the challenge of classifying datasets with narrow velocity and time distribution, SVM classifier is used. The multi-class ``one-vs-all'  Quadratic SVM classifier is used for classifying all the three sets of data, where the classifier functions $f_i(\cdot)$ in~\eqref{eq:SVM-1} are quadratic functions. Our observation is that the classification accuracy for datasets generated using parameters shown in Tables \ref{tb:dataset_l} and \ref{tb:dataset_m} is 100\%. For the dataset generated using parameters shown in Table \ref{tb:dataset_s}, the accuracy obtained is 90\% and the classification is shown in Figure \ref{fig:SVMProcessed4}. The increased accuracy shows that the benefit of Quadratic SVM (QSVM) in providing a  significant performance increase on narrowly distributed datasets when compared to the K-means++ algorithm.

\begin{figure}[h!]
\centering
\captionsetup{justification=centering}
\includegraphics[width=8cm]{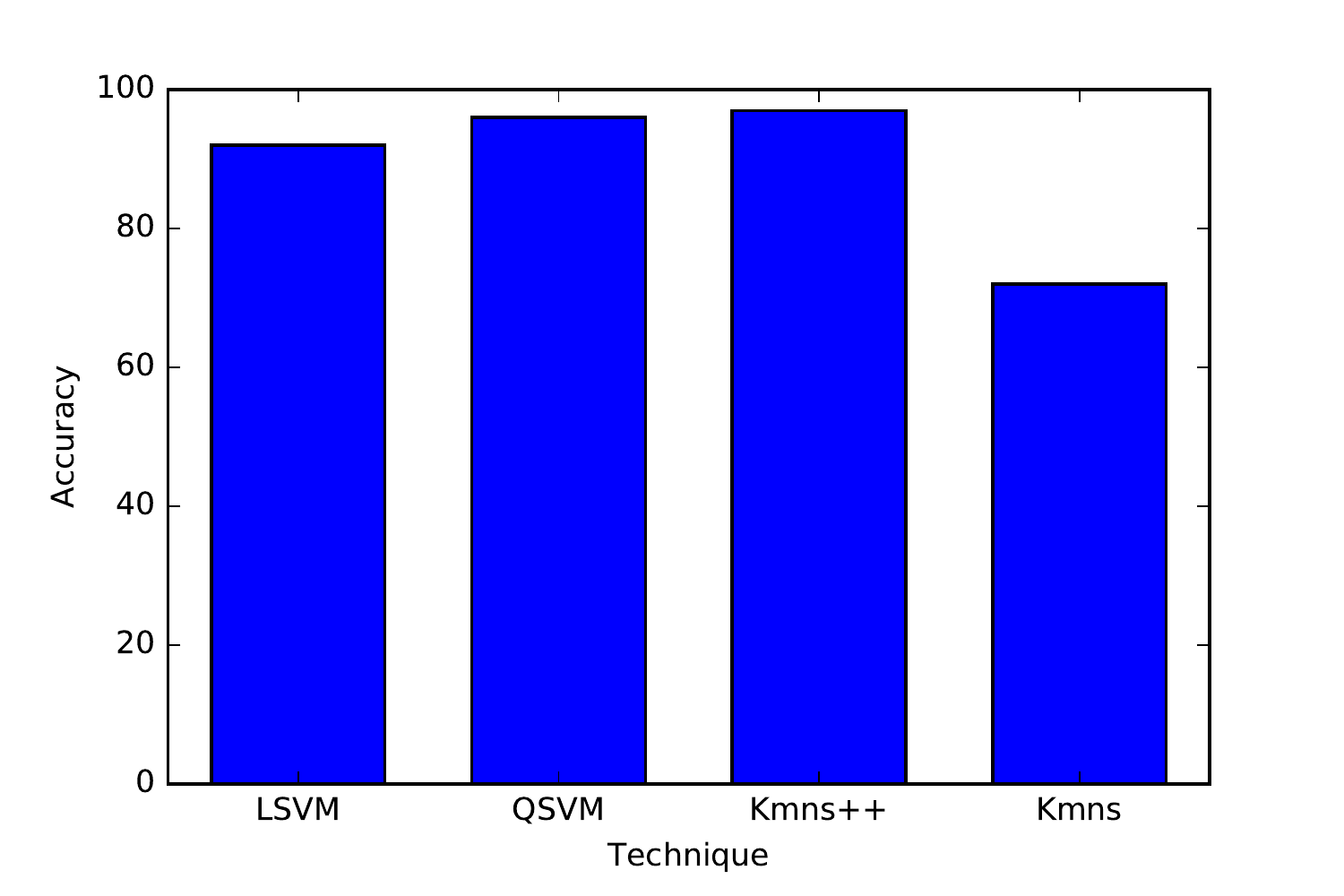}
\caption{Accuracy for processed dataset. LSVM: Linear SVM; QSVM: Quadratic SVM; Kmns$++$: K-means$++$ clustering; Kmns: K-means clustering}
\label{fig:ProcessedAccuracy}
\end{figure}

To compare the performance of different machine learning algorithms, we also plot the accuracy levels of various algorithms. Figure \ref{fig:ProcessedAccuracy} shows the accuracy levels using K-means, K-means++, Quadratic SVM (QSVM), and Linear SVM (LSVM). For K-means algorithm, the accuracy is based on the average obtained using different datasets. For SVM, the accuracy is obtained using a cross validation factor of 10 and then averaged for different datasets. Our observations include: (1) K-means++ provides better performance than the standard K-means; (2) SVM provides similar performance as K-means++ in most case; and (3) QSVM provides most robust performance even when the velocity distributions and initial time instants for all targets are in a small range. 


\section{Conclusion and Future Work}\label{sec:con}
Identifying correct data-target pairing is essential for the situational awareness that the intelligent operation of unmanned systems relies on. This paper studied data pattern recognition for multi-target localization from a list of spatially distributed sensors. In contrast to most existing methods that do not take into consideration the data correlation, we proposed to analyze the data correlation of unlabeled data from different sensors. A three-step approach, i.e., pre-processing, classification, recovering, was proposed, where numerous machine learning algorithms, including both K-means and SVM, were used to provide reliable classification for a variety of target motions.  In addition, simulation studies were provided to show that the proposed method offers a highly accurate solution for recognizing data patterns. Our future work will focus on studying more general road network as well as the consideration of false alarms and missing detections.

\bibliographystyle{IEEEtran}
\bibliography{refs}

\end{document}